\documentclass[aps,prd]{revtex4}
\usepackage{graphicx}   
\def \be{\begin{eqnarray}}
\def \ee{\end{eqnarray}}

\begin{document}
\title{Partial wave interference mechanism in gluonic 
dissociation of $J/\psi$}
\vskip 0.2in
\author{ Binoy K.\ \surname{Patra}$^1$ and V. J. Menon$^2$}
\affiliation{$^1$ Dept. of Physics, Indian Institute of Technology,
Roorkee 247 667, India\\
$^2$ Dept. of Physics, Banaras Hindu University, Varanasi 221 005, India}

\begin{abstract}
We explicitly take into account the effect of hydrodynamic expansion
profile on the gluonic breakup of $J/\psi$'s produced in an equilibrating
parton plasma. Attention is paid to the space-time inhomogeneities as well
as Lorentz frames while deriving new expressions for the gluon number
density $n_g$, average dissociation rate $\langle \tilde{\Gamma} \rangle$,
and $\psi$ survival probability $S$. A novel type of partial
wave {\em interference} mechanism is found to operate in the formula
of $\langle \tilde{\Gamma} \rangle$.
Nonrelativistic longitudinal expansion fro small length of the initial
cylinder is found to push the $S(p_T)$ graph above the no flow case
considered by us earlier~\cite{rev1}. However, relativistic flow corresponding
to large length of the initial cylinder pushes the curve of $S(p_T)$
downwards at LHC but upwards at RHIC. This mutually different effect
on $S(p_T)$ may be attributed to the different initial temperatures generated
at LHC and RHIC.
\end{abstract}


\maketitle

\section*{Introduction}
Among the well known mechanisms of $J/\psi$ dissociation~\cite{screen}-\cite{xu} the one due to
gluonic bombardment deserves special attention here. Recently the
present authors~\cite{rev1} considered the statistical mechanics of important
physical observables {\em viz.} the gluon number density, thermally-averaged
$g-\psi$ break-up rate, and the $\psi$ meson survival probability
appropriate to RHIC/LHC initial conditions. It is a well-recognized fact 
that the longitudinal/transverse
expansion of the medium controls the master rate equations~\cite{biro}
for the time-evolution of the plasma temperature and parton fugacities. But 
the literature
{\it does not} tell how the fluid velocity profile itself influences the
Lorentz transformations connecting the rest frames of the fireball, plasma,
and $\psi$ meson. In other words, since the {\it flow velocity profile} causes 
inhomogeneities in space-time, hence the scenario
of $J/\psi$ gluonic break-up may be affected in a quite nontrivial manner
and the aim of the present paper is to address this hitherto unsolved
problem.

\section*{Theory and Calculations}
If $K=(K^0,\vec{K})$ is the the gluon $4$-momentum in the fireball 
rest frame and $k =(k^0,\vec{k})$, the gluon $4$-momentum in the local 
comoving frame of the plasma is then given by Lorentz transformations
\be
K\cdot u = k^0~;~ K^0/k^0 = \gamma (1 +\vec{v}\cdot \hat{k}),
\ee
where $u=(\gamma, \gamma \vec{v})$ is the fluid $4$ velocity. Then the gluon 
number density using the Bose-Einstein distribution can be expressed as 
\be
n_g(x) =\frac{16}{\pi^2} \gamma T^3 \sum_{n=1}^\infty \frac{\lambda_g^n}{n^3}
\ee
This result shows in a compact 
manner how the
number density depends upon $\gamma$, $T$, and $\lambda_g$. Next, we turn 
to the question of applying statistical mechanics to
gluonic break-up of the $J/\psi$ moving inside an expanding parton
plasma. In the fireball frame consider a $\psi$ meson of mass
$m_\psi$, four momentum $p_\psi = (p_\psi^0, \vec{p}_\psi)$, three
velocity $\vec{v}_\psi$ and dilation factor $\gamma_\psi$.
The invariant quantum mechanical dissociation rate
$\Gamma$ for $g-\psi$ collision may be written as
\be
\Gamma = v_{{}_{\rm{rel}}} \sigma 
\ee
where $v_{\rm{rel}}$ is the relative flux and $\sigma$ the cross section
measured in any chosen frame. Its  thermal average over gluon
momentum in the {\it fireball} frame reads
\be
\langle \Gamma (x) \rangle = \frac{16}{n_g(x)} \int \frac{d^3K}{{(2\pi)}^3}~
\Gamma f
\ee
Let $q=(q^0,\vec{q})$ be the gluon $4$ momentum measured in $\psi$
meson {\it rest} frame. Since the relative flux becomes
$v_{\rm{rel}}^{\rm{Rest}} = c= 1$ hence our invariant $\Gamma$
reduces to the QCD~\cite{BP} based cross section
\be
\Gamma = \sigma_{{}_{\rm{Rest}}} = B {(Q^0-1)}^{3/2}/{Q^0}^5~;~ 
q^0 \ge \epsilon_\psi \nonumber\\
Q^0 =\frac{q^0}{\epsilon_\psi} \ge 1~;~ B=\frac{2\pi}{3}{\left(\frac{32}{3}
\right)}^2
\frac{1}{m_c {(\epsilon_\psi m_c)}^{1/2}}
\ee
where $\epsilon_\psi$ is the $J/\psi$ binding energy and $m_c$ the
charmed quark mass. The energy variable for the massless gluon 
transforms {\em via}
\be
K^0 = \gamma_\psi \left( q^0 +\vec{v}_\psi \cdot \vec{q}~\right)
= \gamma_\psi q^0 \left( 1 + |\vec{v}_\psi| \cos \theta_{q \psi} \right)
\ee
with $\theta_{q \psi}$ being the angle between $\hat{q}$ and $\hat{v}_\psi$
unit vectors. Furthermore, the fluid $4$- velocity $w = (w^0,\vec{w})$ seen
in $\psi$ rest frame will be given by the Lorentz transformations
\be
w^0 &=& \gamma_\psi \gamma \left( 1 - \vec{v} \cdot \vec{v}_\psi 
\right) \nonumber\\
\vec{w} &=& \gamma \left[ \vec{v}  - \gamma_\psi \vec{v}_\psi + (\gamma_\psi -1)
(\vec{v} \cdot \hat{v}_\psi) \hat{v}_\psi \right]\quad ,
\ee
and the scalar product
\be
K\cdot u = q \cdot w = q^0 w^0 - q^0 |\vec{w}| \cos \theta_{qw}
\ee
where $\theta_{qw}$ is the angle between $\hat{q}$ and $\hat{w}$. Finally, the 
thermally-averaged rate of (4) can be calculated as~\cite{euro2}
\be
\langle \Gamma(x) \rangle = \frac{8 \epsilon_\psi^3 \gamma_\psi}{\pi^2 n_g} 
\sum_{n=1}^\infty \lambda_g^n \int_1^\infty
dQ^0 {Q^0}^2 \sigma_{\rm{Rest}} (Q^0) e^{-C_n Q^0} \nonumber\\
\times \left[ \frac{}{} I_0 (\rho_n) + I_1 (\rho_n) |\vec{v}_\psi| 
\cos \theta_{\psi w} \frac{}{} \right]
\ee
where $C_n = n \epsilon_\psi w^0/T$, and
$\rho_n =n \epsilon_\psi Q^0 |\vec{w}|/T$. This demonstrate how the mean 
dissociation rate $\langle \Gamma(x) \rangle$ depends on the
{\it hydrodynamic}  flow through $|\vec{w}|$ (or $w^0$) as well as
the angle $\theta_{\psi w}$.\\
From the analytical viewpoint it is much more advisable to work with
the modified rate
\be
\langle \tilde{\Gamma} (x) \rangle &\equiv& n_g(x) \langle \Gamma(x) \rangle
\nonumber\\
&\approx & \frac{8 \epsilon_\psi^3 \gamma_\psi
}{\pi^2}~\lambda_g \int_1^\infty
dQ^0 {Q^0}^2 \sigma_{{}_{\rm{Rest}}}~H \nonumber\\
&\propto& \lambda_g \gamma_\psi H 
\ee
Here the {\em entire} dependence on the flow velocity $w$ is contained in the
function
\be 
H \equiv e^{-C_1 Q^0_p} \left[ \frac{}{} I_0(D_1 Q^0_p)
+I_1 (D_1 Q^0_p) \mid \vec{v}_\psi \mid \cos \theta_{\psi w} 
\frac{}{} \right] ~, D_1 \equiv \frac{\epsilon_\psi |\vec{w}|}{T}
\ee
 We
are now ready to discuss some consequences of (10) in {\em three} cases
{\em viz.} static medium in the fireball frame, no flow in the
$J/\psi$ rest frame, and ultrarelativistic flow in either frame.\\

First, we consider static medium in fireball frame where
$\vec{v}=\vec{0}~,~ \gamma=1~, w^0 = \gamma_\psi~;~\vec{w} = -\gamma_\psi 
\vec{v}_\psi$.
This is precisely the case treated in our earlier paper~\cite{rev1}.
Due to the assumed absence of flow there is no inhomogeneity with respect
to $x$.
\begin{figure}
   {\includegraphics[scale=0.3]{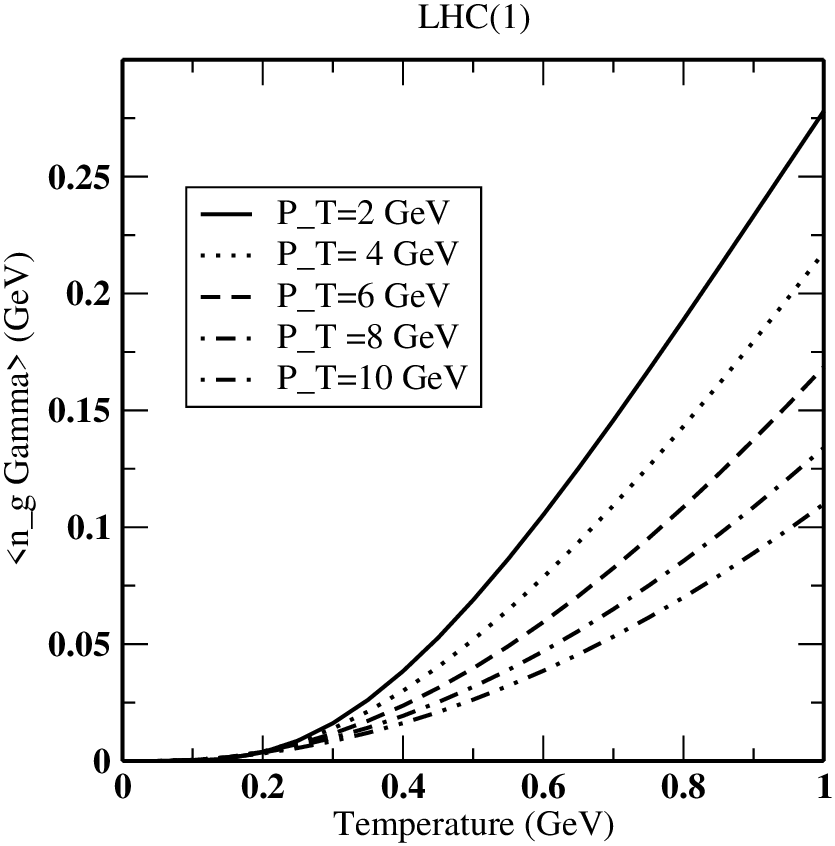}}
\hskip 0.15 in
   {\includegraphics[scale=0.3]{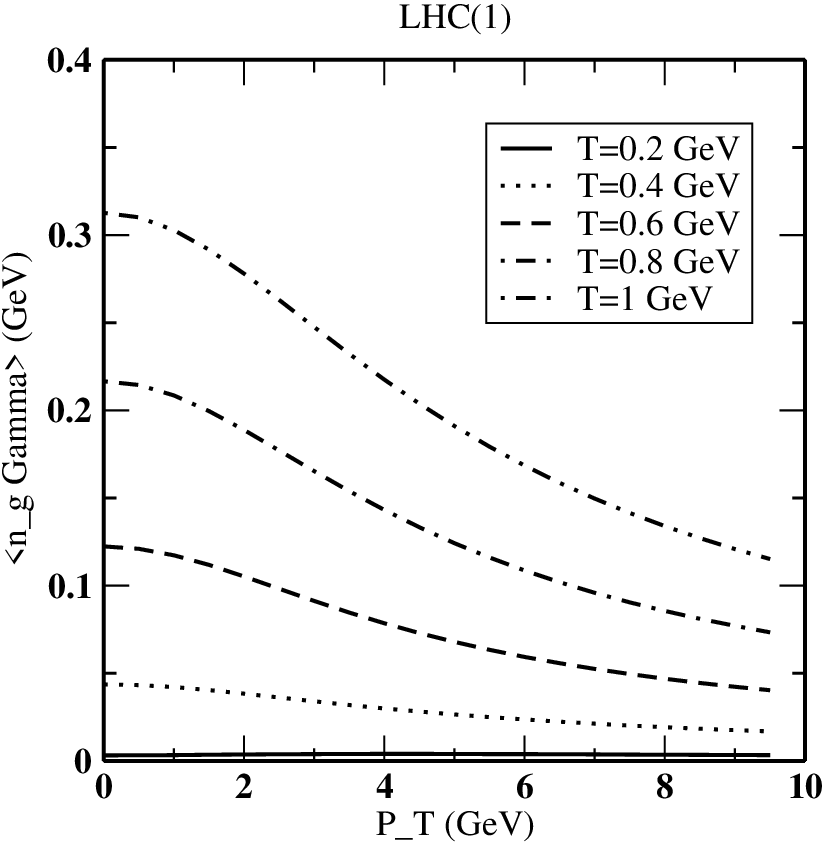}}
\caption{$\langle \tilde{\Gamma}(x) \rangle$ as a function of temperature 
(transverse momenta) at different transverse 
momenta (temperature) in the absence of flow.  }
\end{figure}
At fixed $p_T$ the steady {\em increase} of $\langle \tilde{\Gamma} 
\rangle$ with $T$ in Fig.1 is caused by the growing $\exp{(-C_1^\pm Q^0)}$ 
factors of the estimate(11) whereas at fixed $T$ the monotonic {\em decrease} 
of $\langle \tilde{\Gamma} \rangle$ with $p_T$ in Fig.1 has a very
interesting explanation. For the case under study $\vec{w} =
-\gamma_\psi \vec{v}_\psi$ is antiparallel to $\vec{v}_\psi$ so that
$\cos \theta_{\psi w} =-1$. Hence partial wave terms $I_0$ and $I_1$ of
$H$ {\em interfere} destructively in Fig. 4 making
$\langle \tilde{\Gamma} \rangle$  small as $\mid \vec{v}_\psi \mid$
grows.\\

Secondly, we consider no flow in $J/\psi$ rest frame where
$3$ velocities of the plasma and $\psi$ meson coincide at some $x$
in the fireball frame,i.e., $\vec{v} = \vec{v}_\psi$. Here the values
are consistently higher than the above no-flow case.\\

Thirdly, we illustrate the case of 
both the $J/\psi$ and plasma moving ultrarelativistically (in the tranverse
and longitudinal directions, respectively) with
$\vec{v}=0.9~\hat{e}_z$. The rough estimate (10) becomes
\be
\langle \tilde{\Gamma}(x) \rangle  \propto \frac{\lambda_g T}{\gamma}
\exp \left( -\frac{\epsilon_\psi Q^0_p}{2T \gamma \gamma_\psi} \right)
\left[ 1 - \frac{{|\vec{v}_\psi|}^2}{\gamma} \right]
\ee
\begin{figure}
   {\includegraphics[scale=0.3]{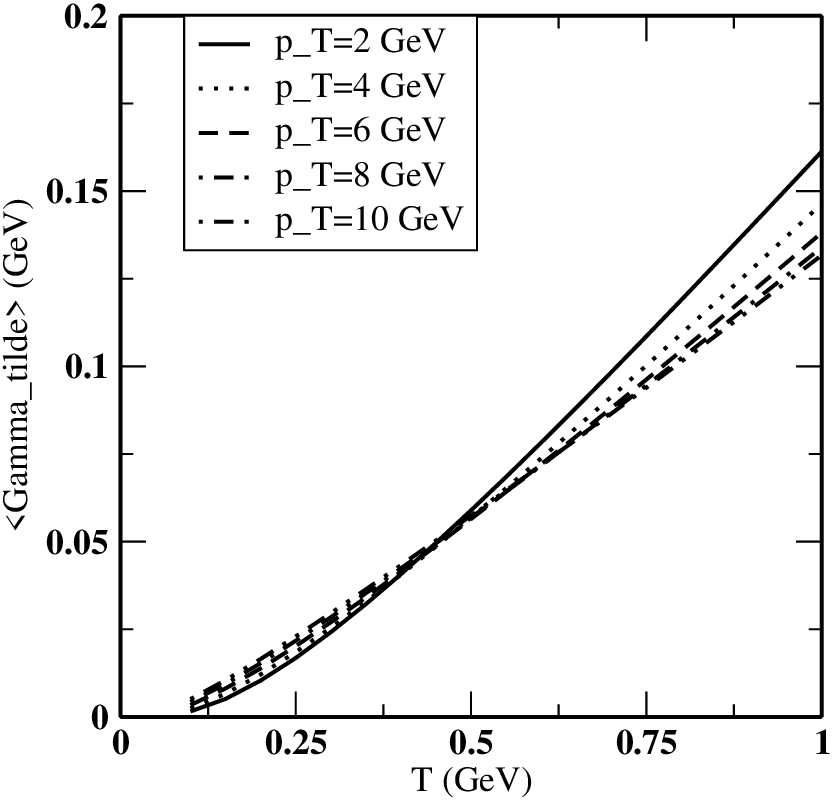}}
\hskip 0.1 in
   {\includegraphics[scale=0.3]{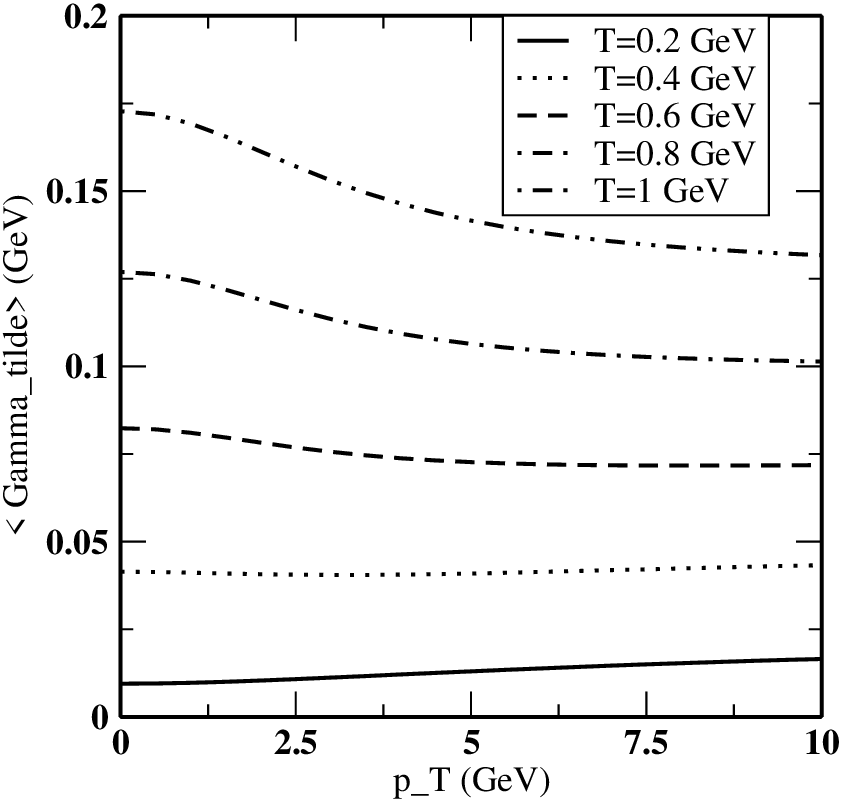}}
\caption{$\langle \tilde{\Gamma}(x) \rangle$ using Eqs.(9,10)
as a function of temperature (transverse momenta) at different transverse 
momenta (temperature) for the ultrarelativistic longitudinal flow 
velocity $v=0.9~c$.  }
\end{figure}
At fixed $p_T$, $v$ the exponential in (11) tend to $0$ as
$T \rightarrow 0$ and tends to $1$ as $T \rightarrow \infty$. 
Therefore, the growing trend
of $\langle \tilde{\Gamma}(x) \rangle$ with $T$ in Fig.2 is 
understandable whereas at fixed $T$, $v$ the rich behaviour of 
$\langle \tilde{\Gamma}(x) \rangle$ with $p_T$ in Fig.2 arises from a
sensitive competition between the bracketed factors of (11). In fact, at 
lower temperatures $T,\le 0.4$ GeV the exponential
factor increases dominantly with $p_T$ causing
$\langle \tilde{\Gamma}(x) \rangle$ to grow, but at higher temperatures 
$T \ge 0.8 $ GeV the third bracket in (12) decreases prominantly with 
$p_T$ causing $\langle \tilde{\Gamma} \rangle$ to drop. In the case of pure
transverse expansion of the plasma $\cos \theta_{\psi w}$ can even become
$+1$, implying constructive interference between $I_0$ and $I_1$ 
in (11)~\cite{euro3}.

\section*{$J/\psi$ SURVIVAL PROBABILITY}
Suppose at general instant $t$ in the 
fireball frame the plasma is contained inside a {\it cylinder} of radius $R$,
and expanding longitudinally with a velocity 
\be
\vec{v} = z\hat{e}_z/t~;~ -L/2 \le z \le +L/2.
\ee
Then the effective
survival chance of a chosen $\psi$ meson will be given by the exponential
$e^{-W}$ with
\be
W = \int_{t_I}^{t_{II}} dt~\langle \tilde{\Gamma}[t] \rangle
\ee
where $t_I = t_i +\gamma_\psi \tau_F$ and $t_{II} ={\rm{min}} (t_I+t_{RI},
t_{\rm{life}})$ with $t_{RI}$ time-taken by $J/\psi$ to traverse the system.
The time dependence of gluon-number density have been provided by the
solution of the master-rate equation of a chemically evolving
plasma~\cite{biro}.\\

Upon averaging $e^{-W}$ over the production configuration of the $\psi$'s
we arrive at the final expression for the net survival probability
\be
S(p_T) &=& \int_{V_I} d^3 x_\psi^I (R_I^2 -{r_\psi^I}^2 ) e^{-W}/
\int_{V_I} d^3 x_\psi^I (R_I^2 -{r_\psi^I}^2 ) \nonumber\\
d^3 x_\psi^I &=& d r_\psi^I~r_\psi^I~d \phi_\psi^I~d z_\psi^I
\ee
\begin{figure}[!tbh]
   {\includegraphics[scale=0.3]{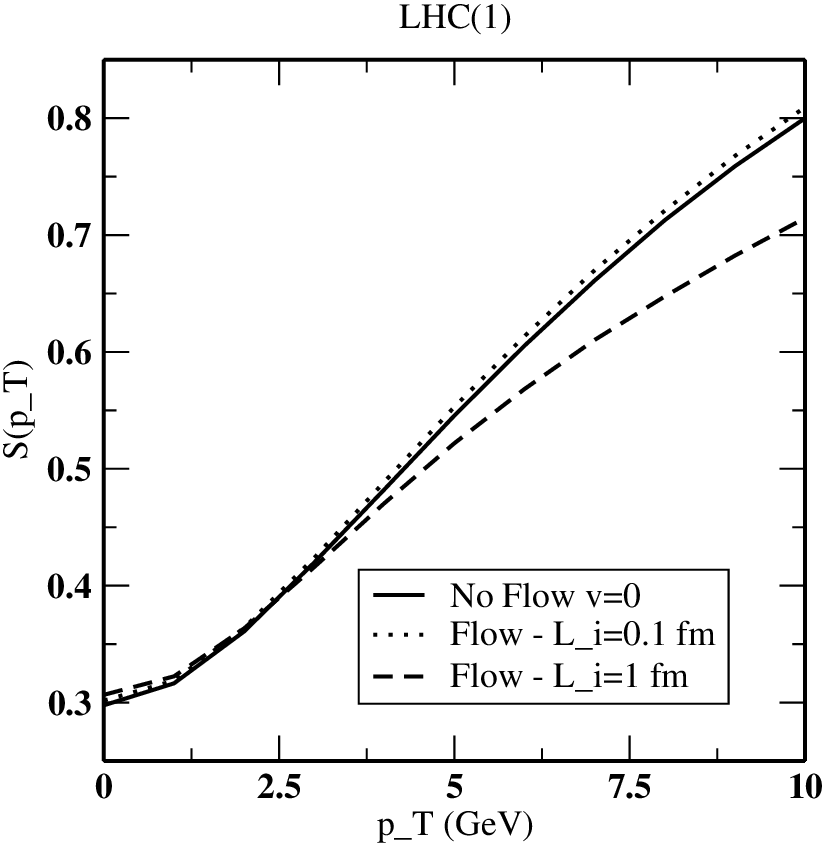}}
\hskip 0.1 in
   {\includegraphics[scale=0.3]{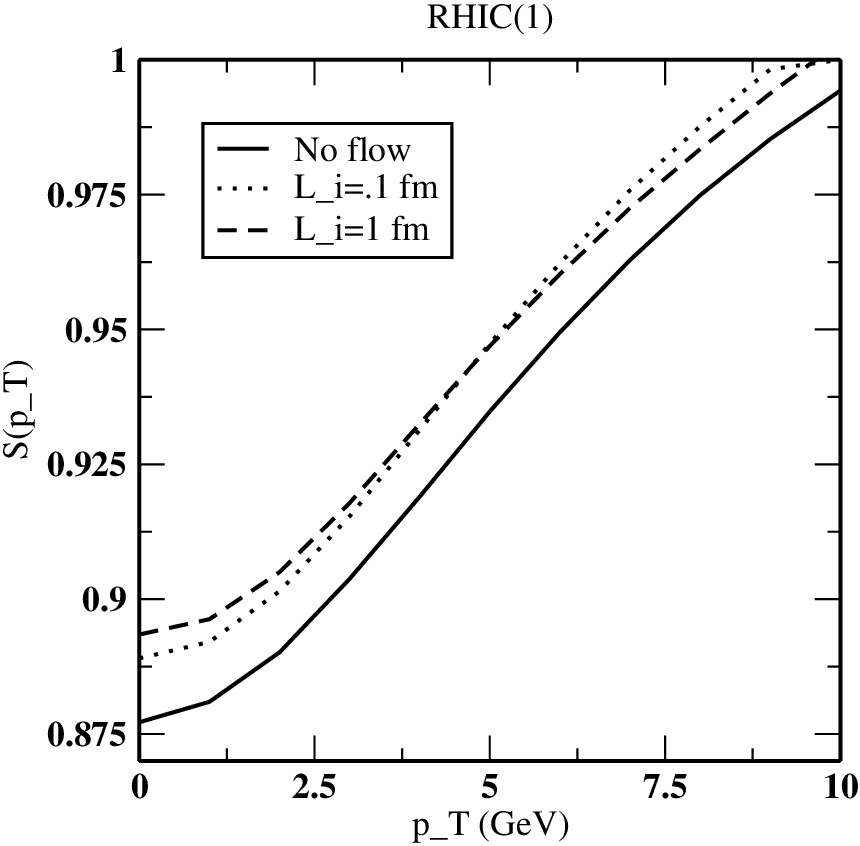}}
\caption{ The solid curve is the result of Ref.~\protect \cite{rev1}, 
i.e., in the {\em absence of flow}
while the dotted and dashed curves represent 
when the plasma is undergoing longitudinal expansion
with the initial values of the length of the cylinder $L_i=0.1$ fm and
$1$ fm, respectively.}
\end{figure}
For chosen creation configuration (label I)
of the $\psi$ meson the function $W$ was first computed from (14) and then 
the survival probability was numerically evaluated using (15). Fig.3 
show the dependence of $S(p_T)$ on the transverse momentum corresponding to the
LHC(1) and RHIC(1) initial conditions~\cite{hijing}.
The dotted $(L_i=0.1$ fm) and dashed ($L_i=1$ fm) curves are computed in the
{\em presence} of longitudinal flow while the solid curve is borrowed from
~\cite{rev1} in the {\em absence} of hydrodynamic flow profile. 
\section*{Discussions and Summary}
Nonrelativistic longitudinal expansion fro small length of the initial
cylinder is found to push the $S(p_T)$ graph above the no flow case
considered by us earlier~\cite{rev1}. However, relativistic flow corresponding
to large length of the initial cylinder pushes the curve of $S(p_T)$
downwards at LHC but upwards at RHIC. This mutually different effect
on $S(p_T)$ may be attributed to the constructive/destructive 
interference in (11) due to different initial conditions
at LHC and RHIC.

\end{document}